\documentclass{svjour3}                    % onecolumn (standard format)
\smartqed  % flush right qed marks, e.g. at end of proof
\usepackage{graphicx}
\usepackage{fix-cm}
\journalname{Journal of Grid Computing}
\begin{document}

\title{Development of Grid e-Infrastructure in South-Eastern Europe}

\subtitle{}

%\titlerunning{Short form of title}        % if too long for running head

\author{Antun Bala\v z\and
		Ognjen Prnjat\and
		Du\v san Vudragovi\' c\and
		Vladimir Slavni\' c\and
		Ioannis Liabotis\and
		Emanouil Atanassov\and
		Boro Jakimovski\and
		Mihajlo Savi\' c
}

\authorrunning{A. Bala\v z, O. Prnjat, D. Vudragovi\' c et al.} 

\institute{Antun Bala\v z\at
           Scientific Computing Laboratory, Institute of Physics Belgrade, University of Belgrade, Serbia,
           \email{antun@ipb.ac.rs}
           \and
           Ognjen Prnjat\at
           Greek Research and Technology Network - GRNET, Greece
            \and
           Dusan Vudragovi\' c and Vladimir Slavni\' c\at
           Scientific Computing Laboratory, Institute of Physics Belgrade, University of Belgrade, Serbia
           \and
           Ioannis Liabotis\at
           Greek Research and technology Network - GRNET, Greece
           \and
           Emanouil Atanassov\at
           Institute for Parallel Processing, Bulgarian Academy of Sciences
           \and
           Boro Jakimovski\at
           SS. Cyril and Methodius University in Skopje, FYR of Macedonia
           \and
           Mihajlo Savi\' c\at
           University of Banja Luka, Bosnia and Herzegovina
}

\date{}
% The correct dates will be entered by the editor

\maketitle

\begin{abstract}
Over the period of 6 years and three phases, the SEE-GRID programme has established a strong regional human network in the area of distributed scientific computing and has set up a powerful regional Grid infrastructure. It attracted a number of user communities and applications from diverse fields from countries throughout the South-Eastern Europe. From the infrastructure point view, the first project phase has established a pilot Grid infrastructure with more than 20 resource centers in 11 countries. During the subsequent two phases of the project, the infrastructure has grown to currently 55 resource centers with more than 6600 CPUs and 750 TBs of disk storage, distributed in 16 participating countries. Inclusion of new resource centers to the existing infrastructure, as well as a support to new user communities, has demanded setup of regionally distributed core services, development of new monitoring and operational tools, and close collaboration of all partner institution in managing such a complex infrastructure. In this paper we give an overview of the development and current status of SEE-GRID regional infrastructure and describe its transition to the NGI-based Grid model in EGI, with the strong SEE regional collaboration.

\end{abstract}

%----------------------------------------------------------------------------------------------------%

\section{Introduction}
\label{intro}

The transition of the traditional science to e-Science is fueled by the ever increasing need for processing of exceedingly large amounts of data and exponentially increasing computational requirements: in order to realistically describe and solve real-world problems, numerical simulations are becoming more detailed, experimental sciences use more sophisticated sensors to make precise measurements; and shift from the individuals-based science work towards collaborative research model now starts to dominate.

Computing resources and services able to support needs of such a new model of scientific work are available at different layers: local computing centers, national and regional computing centers, and supercomputing centers. The gap between the needs of various user communities and dispersed computing resources able to satisfy their requirements is effectively bridged by introduction of Grid technology on the top of the networking layer and local resource management layers.

Computing Grids are conceptually not unlike electrical grids. In an electrical grid, the wall outlets allow us to link to and use an infrastructure of resources, which generate, distribute, and bill for electrical power. When we connect to the electrical grid, we do not need to know details on the power plant currently generating the electricity we use. In the same way Grid technology uses middleware layer to coordinate and organize into one logical resource a set of available distributed computing and storage resources across a network, allowing users to access them in a unified fashion. The computing Grids, like electrical grids, aim to provide users with easy access to all the resources they need, whenever they need them, regardless of the underlying physical topology and management model of individual clusters.

Grids address two distinct but related goals: providing remote access to information technology (IT) assets, and aggregating processing and storage power. The most obvious resources included in Grids are processors (CPUs) and data storage systems, but Grids also can encompass various sensors, applications, and other advanced types of resources. One of the first commonly known Grid initiatives was the SETI@HOME project, which solicited several millions of volunteers to download a screensaver, which was able to use idle processor time to analyze the astronomical data in the search for extraterrestrial life.

In the past 6 years the European Commission has funded, through a number of targeted initiatives, activation of new user communities and enabling collaborative research across a number of fields in order to close existing technological and scientific gaps. In addition, this helps in bridging the digital divide, stimulating research and consequently alleviating the brain drain in the less-developed regions of Europe. This was especially successful in the South-Eastern Europe (SEE), where a number of such initiatives show excellent results. In the Grid arena, the South-East European GRid eInfrastructure Development (SEE-GRID) series of projects~\cite{see-grid,see-grid-2}, through its first two 2-year phases, has established a strong human network in the area of scientific computing and has set up a powerful regional Grid infrastructure, attracting large number of applications from diverse fields from countries throughout the South-Eastern Europe. The third 2-year phase of the SEE-GRID programme, SEE-GRID-SCI ~\cite{see-grid-sci} project, has aimed and succeeded in having a catalytic effect on a number of SEE user groups, with a strong focus on the key seismological, meteorological, and environmental communities.

One of the main successes of the SEE-GRID programme is cumulative structuring effort on the establishment of National Grid Initiatives (NGIs) in SEE countries and collaborative work on achieving sustainable model of operation, supported strongly from national funding sources. The regional SEE-GRID initiative has also supported and coordinated a successful transition of all SEE countries from the centralized operations model to the NGI-based EGI infrastructure, which is clearly visible from the participation of all partner countries in the 4-year EGI-InSPIRE project~\cite{egi}. 

%----------------------------------------------------------------------------------------------------%

\section{Resource Centers}
\label{rc}
The regional Grid infrastructure operated by SEE-GRID-SCI project was built on top of the pilot infrastructure established by the first SEE-GRID project (2004-2006), which was since then substantially extended and enlarged in terms of resources and number of Grid sites, and upgraded in terms of the deployed middleware and core services provided to existing and new user communities during the SEE-GRID-2 project (2006-2008).

The operations activity adopted the pragmatic model of the 2-layered infrastructures in which mature sites were migrated to the EGEE~\cite{egee} production infrastructure, while the start-up sites from new institutes and user communities were incubated within the SEE-GRID infrastructure until they were ready to follow the requirements of the full-scale production infrastructure. In this way, both SEE-wide and national-level applications were able to benefit from the computing resources of both infrastructures, by mainly using the pilot infrastructure in the incubation phase and production infrastructure later, when they reach the production phase. Moreover, this approach ensured that smaller sites, typical for the region, have a chance to be a part of the regional SEE-GRID infrastructure acting as an incubator for their maturing into EGEE production.

As applications developed in the region have matured, new Virtual Organizations (VOs) have spun off with the relevant core services supported by the SEE-GRID-SCI operations activity SA1. Discipline-specific services were deployed in multiple instances (for failover and for achieving load-balancing through a wide geographic distribution) over the e-Infrastructure and operationally maintained and supported by SA1. Sophisticated operational tools, some of them being developed within the joint research activity JRA1 of the SEE-GRID-SCI project, were used to enhance infrastructure performance.

SEE-GRID-SCI project has continued to operate and further extend, develop and improve this infrastructure, with the aim to cater for the needs of all activated user communities in the region, with special emphasis on the three identified target areas: meteorology, seismology, and environmental sciences. Apart from computing and storage resources made available to these user communities, SA1 activity provided and maintained a set of existing and new operational and monitoring tools so as to ensure proper operation of the infrastructure, and a set of primary and secondary core services for all deployed VOs in order to ensure optimal geographical distribution according to the underlying network structure, load sharing, and quality of the service to end users.

\begin{figure}[!t]
  \includegraphics[width=0.97\textwidth]{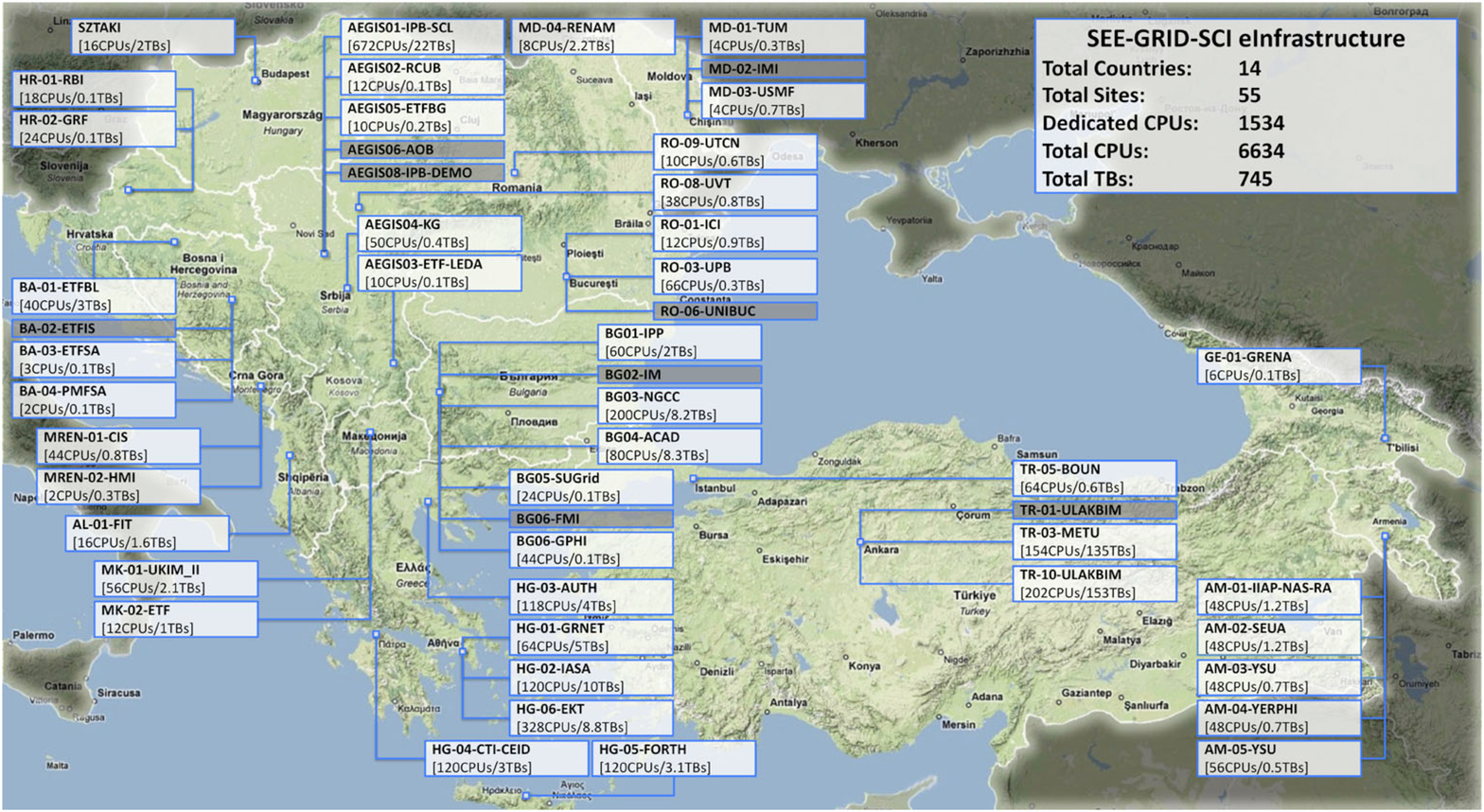}
\caption{Overview of the SEE-GRID-SCI infrastructure.}
\label{inframap}
\end{figure}

Currently SEE-GRID-SCI infrastructure encompasses approximately 55 Grid sites, more than 6600 CPUs, and around 750 TBs of available data storage capacity, which is illustrated in Fig.~\ref{inframap}, with further details given in Table~\ref{infratab}. Overall number of CPUs has grown from 2400 at the beginning of the SEE-GRID-SCI project in May 2008 to currently more than 6600, while the number of dedicated CPUs for SEE-GRID-SCI VOs is currently around 1500. Grid operations activity successfully maintains such a large, geographically disperse and ever-growing infrastructure, harmonizing its operation with the pan-European EGEE/EGI infrastructure. In addition to this, one of the most important achievements of SA1 activity is transfer of knowledge and Grid know-how to all participating countries, and support to their NGI operation teams to reach the level of expertise needed for sustainable NGI-based operational model in EGI.

\begin{table}[!b]
\caption{SEE-GRID-SCI computing and storage resources.}
\label{infratab}
\begin{tabular}{lll}
\hline\noalign{\smallskip}
Country & Total number of CPUs & Total storage [TB] \\
\noalign{\smallskip}\hline\noalign{\smallskip}
Greece 					& 1200  	& 66.8 \\
Bulgaria 				& 1210  	& 42.3 \\
Romania 				& 120  		& 4.0 \\
Turkey 					& 2380  	& 528.0 \\
Hungary 				& 8  		& 2.0 \\
Albania 				& 34  		& 1.3 \\
Bosnia-Herzegovina 	& 80  		& 1.1 \\
FYR of Macedonia 		& 80  		& 4.1 \\
Serbia 					& 974  		& 97.0 \\
Montenegro 				& 40  		& 0.6 \\
Moldova 				& 24  		& 6.5 \\
Croatia 				& 44  		& 0.2 \\
Armenia 				& 424  		& 0.2 \\
Georgia 				& 16  		& 0.1 \\
\noalign{\smallskip}\hline\noalign{\smallskip}
Total 					& 6634  	& 754.2 \\
\noalign{\smallskip}\hline
\end{tabular}
\end{table}

After the completion of the SEE-GRID-SCI project in April 2010, the regional Grid infrastructure was seamlessly integrated to the EGI infrastructure, and continues to support all deployed Virtual Organizations (VOs) and applications developed during the 6-year SEE-GRID programme. The strong human network remains in place and still supports on-going transition of all countries to independent NGI operations through the SEE Regional Operations Centre. The catch-all SEE-GRID Certification Authority will continue its operation until all the countries from the region deploy their own national certification authorities. In terms of Grid operations, currently almost all (with only a few exceptions) NGI operations teams and infrastructures are fully validated by EGI teams, while validation for the remaining SEE countries is expected to finish within a few months, i.e. by mid-2011.

%----------------------------------------------------------------------------------------------------%

\section{User Communities}

The core objective of the SEE-GRID-SCI project was to engage user communities from different regional countries in close collaboration. This strategy had a structuring effect for crucial regional communities. The target applications were selected from core earth science disciplines in the region, namely, seismology, meteorology and environmental protection. Thus, the focus of the project was to engage these three core cross-border communities in the research fields crucial for the region, structured in the form of Virtual Organizations (VO):
\begin{itemize}
\item
Seismology VO had six applications \cite{6,7,8,9,10,11,12} ranging from Seismic Data Service to Earthquake Location Finding, from Numerical Modelling of Mantle Convection to Seismic Risk Assessment. 
\item
Meteorology VO, with two large-scale applications \cite{13,14,15,16,17,18,19,20,21}, follows an innovative approach to weather forecasting that uses a multitude of weather models and bases the final forecast on an ensemble of weather model outputs. The other problem tackled within this VO is the reproduction/forecasting of the airflow over complex terrain.
\item
Environmental Protection VO supports eight applications \cite{22,23,24,25,26,27,28,29,30,31,32} focusing on environmental protection/response and environment-oriented satellite image processing.
\end{itemize}

In the Seismology VO, the work was organized around the development of Seismic Data Server (SDS) application services, providing distributed storage and serving of seismic data from different partner countries, logical organization and indexing of distributed seismic data, and programming tools (called iterators) that provide easy access to seismic data. In terms of applications, the focus was on gridification of five seismology applications from different South-eastern European countries: Seismic Risk Assessment (SRA), Numerical Modeling of Mantle Convection (NMMC3D), Fault Plane Solution (FPS), Earthquake Location Finding (ELF) and Massive Digital Seismological Signal Processing with the Wavelet Analysis (MDSSP-WA).

In the Meteorology VO, with the aim to contribute to the improvement of the forecasts in the Mediterranean, among other techniques, the regional ensemble forecasting technique has been explored in the frame of the SEE-GRID-SCI. Indeed the regional ensemble forecasting system built over the Mediterranean, involves the need of large infrastructure that was not easily available at medium-scale research centres and institutions. For that reason, the Grid infrastructure was explored for its ability to support the high CPU and storage needs of such a regional ensemble forecasting system. This application allowed the meteorological entities participating in the project to assess the probability of a particular weather event to occur. This information is being made freely available (to the participants and to the general public, etc) through the project web page, helping thus when needed, to make the necessary decisions based on this probabilistic information.
In addition, another set of applications permitted the entities participating in the project to improve the quality of the understanding and forecasting of the airflow over regions characterized by the complex terrain. Further an important benefit of this application is the possibility offered to use this model for operational weather forecasting. Operational weather forecasting model chains based on this model have been developed in the frame of this project over Bosnia and Herzegovina, Armenia and Georgia. This is considered as an important benefit for the meteorological services of the aforementioned countries that did not have up to now the infrastructure support to run operationally weather forecasting models for their region.

The Environmental VO has dealt with several important problem areas in the domain of environmental modeling and environmental protection and the applications developed within the VO advanced the scientific knowledge and affected the policy and decision-making process, responding to the EU directives and national priorities. New modeling techniques and algorithms were employed in several of the applications, using the power of the Grid in order to increase the spatial and temporal resolution and obtain more adequate representation of the natural processes under investigation. In other applications, established techniques were used, combined with filters and scripts developed by the project partners in order to accommodate these systems to the specifics of the Balkan region. The beneficiaries of the systems developed during the projectÕs lifetime include not only environmental scientists, but also the relevant governmental and international organizations, for example the international air quality monitoring bodies. By employing the Grid to increase the resolution these applications are now starting to target new beneficiaries like municipal authorities, small and medium enterprises and media. For many of the applications the validation of the models and standardizing the computational processes has been an important achievement, since the methodological aspect of these studies was a challenging one, especially in the Balkan region.

\begin{figure}[!t]
  \includegraphics[width=0.98\textwidth]{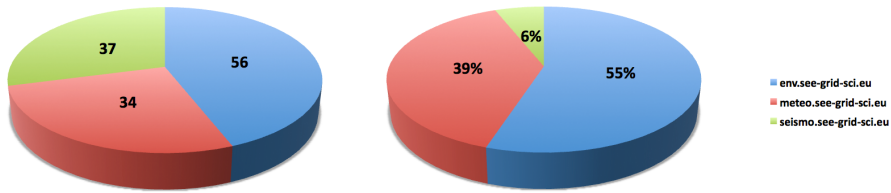}
\caption{The distribution of the size of three target user communities (left, number of end users per VO), and the distribution of the computing resources used by VOs (right).}
\label{users}
\end{figure}

Fig.~\ref{users} gives some details on the size of activated user communities, and the distribution of computing resources they have utilized during the project lifetime. Overall, during the period 2008-2010, SEE-GRID-SCI project has provided more than 22.5 million elapsed CPU hours or 2566 CPU years, and more than 4.5 million jobs were executed on the regional infrastructure. Out of this, SEE-GRID-SCI and national VOs amounted to 16.4 million CPU hours or 1872 CPU years (73\%). The total utilization of dedicated resources (based on the average number of 1050 available CPUs) was quite high, around 89\%, and this has attracted the growth of supported user communities, and enabled them to achieve the enormous amount of new scientific results, as can be seen by the large number of scientific papers published in per-reviewed research journals \cite{33,34,35,36,37,38} and presented at numerous scientific conferences \cite{6,7,8,9,10,11,12,13,14,15,16,17,18,19,20,21,22,23,24,25,26,27,28,29,30,31,32}. The project itself has organized SEE-GRID-SCI User Forum in December 2009, where the most significant results were presented.

\section{Core Services}
\label{cs}

\begin{figure}[!h]
  \includegraphics[width=0.95\textwidth]{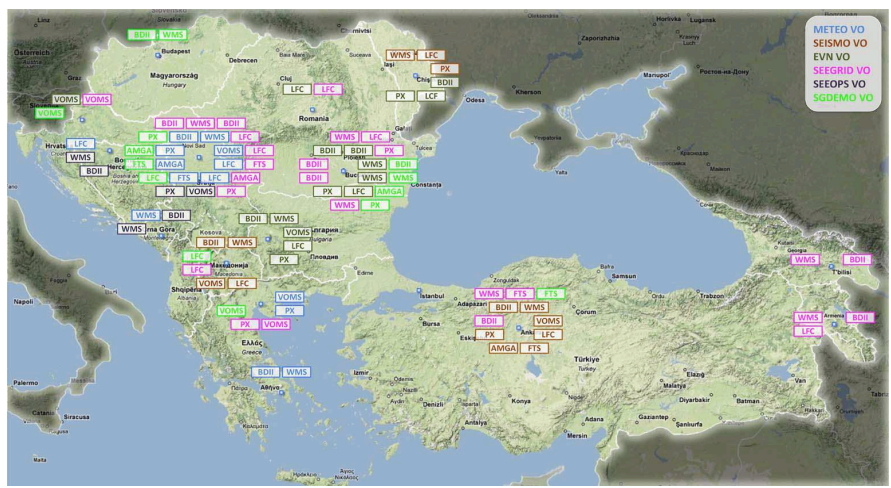}
\caption{Geographical distribution of core services.}
\label{coremap}
\end{figure}

To operationally provide computational and storage resource to the three target scientific communities supported by the SEE-GRID-SCI project, three different VOs have been created: METEO, SEISMO, and ENV VO. The support for these VOs, as well as to the catch-all SEEGRID VO, has been configured on all Resource Centres participating in the regional infrastructure, and a set of core services was installed and deployed by SA1 activity, as illustrated in Fig.~\ref{coremap}.

\begin{table}[!t]
\caption{List of primary core services deployed per VO.}
\label{coretab}
\begin{tabular}{llll}
\hline\noalign{\smallskip}
Service & METEO VO & ENV VO & SEISMO VO \\
\noalign{\smallskip}\hline\noalign{\smallskip}
VOMS	& voms.grid.auth.gr		& voms.ipp.acad.bg				& voms.ulakbim.gov.tr \\
WMS \& LB	& wms.ipb.ac.rs				& wms.ipp.acad.bg				& wms.ulakbim.gov.tr \\
BDII	& bdii.ipb.ac.rs			& bdii.ipp.acad.bg				& bdii.ulakbim.gov.tr \\
LFC		& grid02.rcub.bg.ac.rs		& lfc01.mosigrid.utcluj.ro		& lfc.ulakbim.gov.tr \\
FTS		& grid16.rcub.bg.ac.rs		& 								& fts.ulakbim.gov.tr \\
AMGA	& grid16.rcub.bg.ac.rs		& 								& amga.ulakbim.gov.tr \\
PX		& myproxy.ipb.ac.rs		& myproxy.ipp.acad.bg			& myproxy.ulakbim.gov.tr \\
\noalign{\smallskip}\hline
\end{tabular}
\end{table}

For each VO a primary and secondary VO Management Service (VOMS) has been deployed and maintained by institutes involved in the corresponding VO application development. Additionally, a set of core Grid services was deployed in order to support job management operations (Workload Management System - WMS, Logging and Bookkeeping - LB), Grid information system (Berkeley Database Information Index - BDII), data storage and transfer operations (Logical File Catalog - LFC, File Transfer Service - FTS, ARDA Metadata Grid Application - AMGA), and management of digital credentials (MyProxy - PX). Deployment details of primary core Grid services are given in Table~\ref{coretab}.

%----------------------------------------------------------------------------------------------------%

\section{Grid operations}
\label{ops}

\begin{figure}[!b]
  \includegraphics[width=0.85\textwidth]{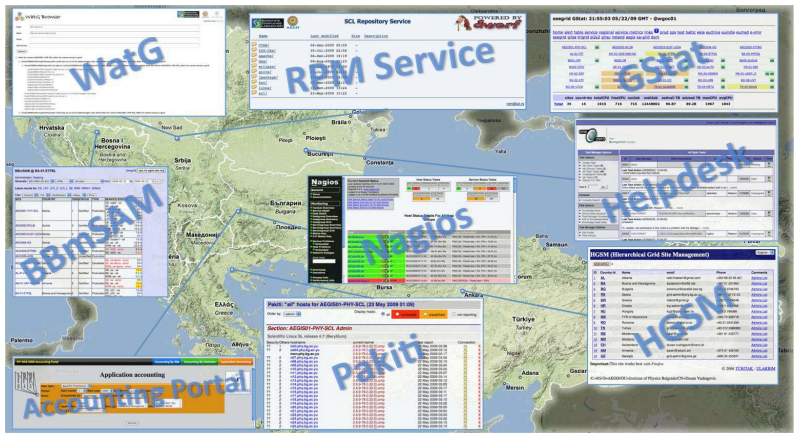}
\caption{Geographical distribution of SEE-GRID operational and monitoring tools.}
\label{toolsmap}
\end{figure}

This section gives brief description of operational procedures and key tools developed during the course of the SEE-GRID programme. In addition, a number of operational tools have been developed, improved and deployed by the SEE-GRID-SCI SA1 activity and used in day-to-day infrastructure management, as illustrated in Fig.~\ref{toolsmap}.
Table~\ref{toolstab} lists all currently deployed tools including those used for monitoring of the infrastructure, some of which are described in more detail in the next section, while Fig.~\ref{toolsmap} gives their geographical distribution, as well as distribution of responsibilities for their deployment and maintaining. The interactions and collaboration on the development and usage of described tools with other Grid initiatives/projects are emphasized wherever applicable.

Recognizing that improvements in the quality and shaping-up of the SEE-GRID infrastructure are an important and continuous effort, necessary for the successful work of SEE-GRID application developers, as well as for the usage of our infrastructure by the existing user communities, the pro-active monitoring of Grid sites in the region was organized through rotating shifts by SA1 country representatives (Grid Infrastructure Managers - GIMs). During each shift, the corresponding GIM is designated as Grid-Operator-On-Duty, or GOOD \cite{goods}.

Basically, the idea is that each GIM (i.e. GIM team from one country) is on duty during one week overseeing the infrastructure and opening trouble tickets in the SEE-GRID Helpdesk to sites from all countries where operational problems are identified using the available monitoring tools. Of course, all GIMs are expected to continually monitor and provide support to sites from their countries - this is their day-to-day duty, in addition to regular regional GOOD shifts. Details of the organization of GOOD shifts are given at the SEE-GRID Wiki \cite{mpi,wiki}. For problems identified by GOODs, trouble tickets were created in the SEE-GRID Helpdesk \cite{helpdesk}, and site managers were expected to deal with such operational problems and provide feedback on the steps taken. Typically, simple problems were resolved within one working day, while for more complex issues typical resolution time was up to three working days.

On the request of applications which need MPI support on sites, GOODs are expected to test MPI setup on all SEE-GRID sites which support MPI. The MPI setup tests are performed at least once a week, and GOODs ensure that the test parallel jobs run at the same time on at least two WNs (to test ssh setup as well). More details can be found on the Wiki page on Testing MPI support~\cite{mpi}.

\begin{table}[!t]
\caption{Deployment of operational and monitoring tools in the SEE-GRID infrastructure.}
\label{toolstab}
\begin{tabular}{ll}
\hline\noalign{\smallskip}
Service & Service URL \\
\noalign{\smallskip}\hline\noalign{\smallskip}
HGSM							& https://hgsm.grid.org.tr/ \\
BBmSAM							& https://c01.grid.etfbl.net/bbmsam/ \\
BBmobileSAM						& https://c01.grid.etfbl.net/bbmsam/mobile.php \\
Gstat							& http://gstat.gridops.org/gstat/seegrid/ \\
Accounting Portal				& http://gserv4.ipp.acad.bg:8080/AccountingPortal/ \\
Nagios							& https://portal.ipp.acad.bg:7443/seegridnagios/ \\
Googlemap						& http://www.grid.org.tr/eng/ \\
MonALISA						& http://monitor.seegrid.grid.pub.ro:8080/ \\
Real Time Monitor				& http://gridportal.hep.ph.ic.ac.uk/rtm/applet.html \\
WatG Browser					& http://watgbrowser.scl.rs:8080/ \\
WMS Monitoring Tool			& http://wmsmon.scl.rs/ \\
Repository Service				& http://rpm.egee-see.org/yum/SEE-GRID/ \\
Dwarf							& https://dwarf.scl.rs/ \\
Grid-Operator-On-Duty			& http://wiki.egee-see.org/index.php/SG\_GOOD \\
Helpdesk						& http://helpdesk.see-grid.eu/ \\
SEE-GRID Wiki					& http://wiki.egee-see.org/index.php/SEE-GRID\_Wiki \\
P-Grade Portal					&http://portal.p-grade.hu/multi-grid \\
\noalign{\smallskip}\hline
\end{tabular}
\end{table}

In this section we describe two selected tools used for Grid operations: HGSM database (used for maintaining the database of Grid resources and personnel), Dwarf portal related to software development and repositories (especially important in maintaining updated of Grid middleware and application software).

\subsection{HGSM}
\label{hgsm}

Hierarchical Grid Site Management - HGSM \cite{hgsm} is a web based management application primarily geared towards Grid site administrators. At the beginning it was designed to store static information about Grid sites and personnel responsible for the sites, but later it evolved to the central information hub, also used for other Grid monitoring and checking services.

\begin{figure}[!t]
  \includegraphics[width=0.95\textwidth]{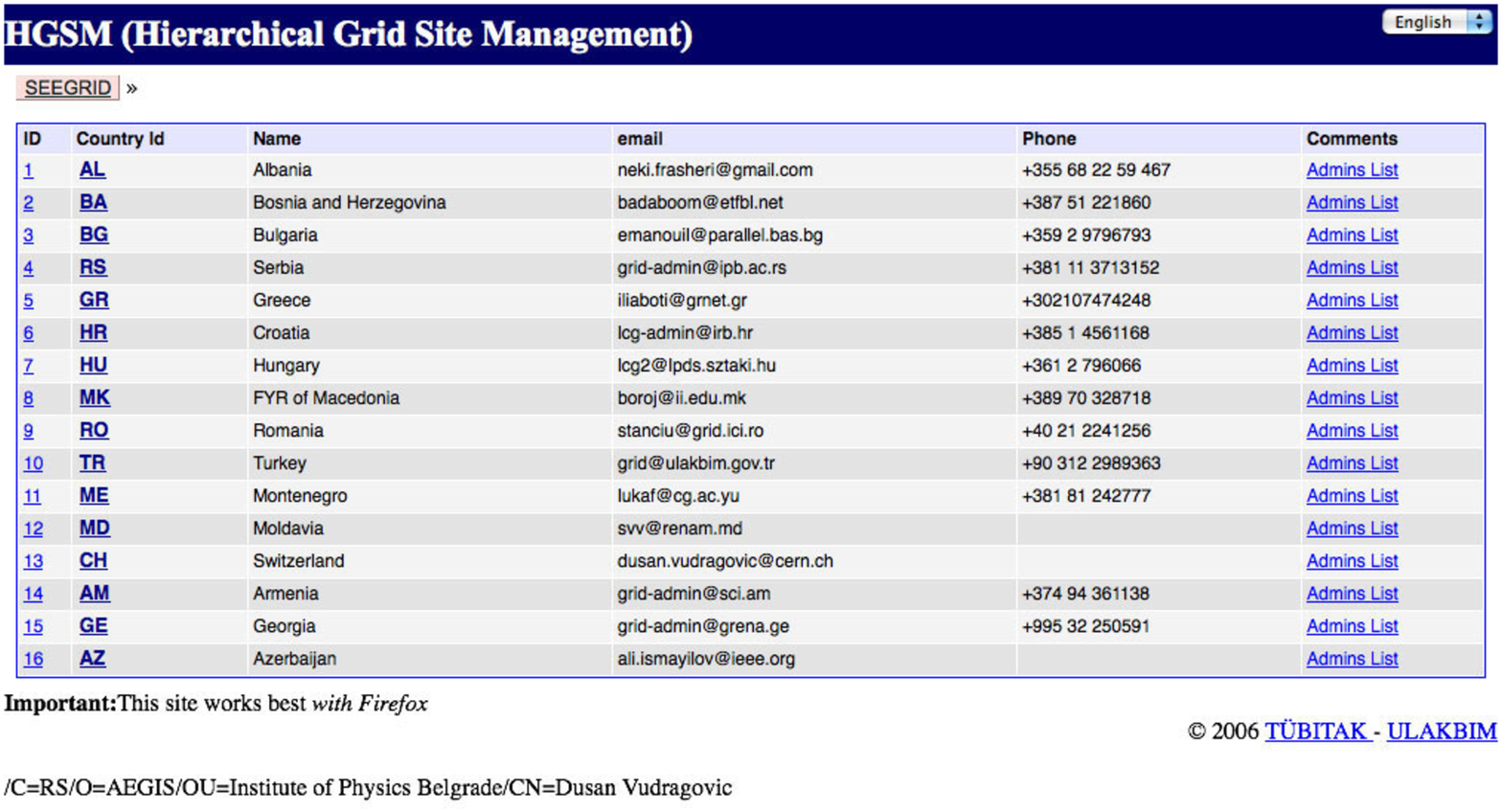}
\caption{Overview of HGSM portal.}
\label{hgmssnap}
\end{figure}

The idea behind the HGSM is to reflect the natural hierarchy present in the infrastructure. For each supported infrastructure, HGSM has a ROC (Regional Operational Centers) associated with it at the top. These ROCs contain the countries that participate in a particular infrastructure. Grid sites of each country are listed under the respective country tree, and all details related to a specific Grid site can be viewed under the respective site entry in the web front end of HGSM, Fig.~\ref{hgmssnap}. The management personnel information is also stored for each organizational level (ROC, country, site), containing contacts with both administrative and management privileges.

While HGSM holds vast information about Grid sites and core services, it also contains personal information for named contacts (names, e-mail addresses and phone numbers). To properly protect this information, HGSM uses a digital certificate-based authentication system. HGSM server only authorizes people with a valid Grid certificate to view the information in HGSM web front-end. Editing information is only allowed to authorized personnel with administrative privileges. The authorization is organized in a hierarchical manner, so that an administrator at the higher level can manage every aspect (including the administrators) at lower organizational levels.

HGSM has already been used by communities and projects other than SEE-GRID, e.g. by the Deployment of Remote Instrumentation Infrastructure - DORII project \cite{dorii}, as well as by the Spanish and Portuguese NGIs \cite{ibergridhgsm}.

%--------------------------------------------------%

\subsection{Dwarf}
\label{dwarf}

\begin{figure}[!b]
  \includegraphics[width=0.98\textwidth]{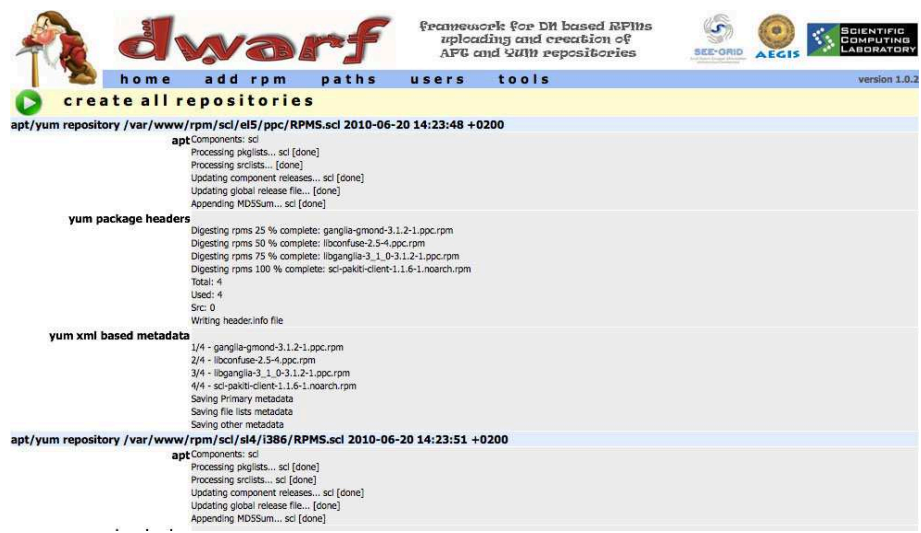}
\caption{Overview of the SEE-GRID Dwarf web portal.}
\label{dwarfsnap}
\end{figure}

Web-based Dwarf tool is composed of the Dwarf web portal~\cite{dwarf}, Dwarf modules and Dwarf database. Using the Public Key Infrastructure (PKI), Dwarf framework provides digital certificate-based management of RPM uploading and creation of APT and YUM repositories. The Dwarf web portal home page, shown in Fig.~\ref{dwarfsnap}, gives an overview of repository structure together with information on the context of each repository, and latest build's timestamp.

From the Dwarf web portal, properly authenticated and authorized user can perform the following operations on the repository:
\begin{itemize}
\item{Create and change repository structure: Users can create paths to new distributions and components, by specifying their names. In the current implementation of the Dwarf framework, the users are able to create APT and YUM repositories, as well as to create a MIRROR to an existing remote repository.}
\item{Package uploading: Users can upload different software packages, but only to sections of the repository for which they are authorized as contributors.}
\item{Build repository: After each RPM upload, a user should re-build the repository structure. If not, Dwarf system will do it automatically, through a cron job.}
\end{itemize}

Dwarf modules are implemented as bash scripts which handle appropriate build actions on various repositories. 

After a new APT repository structure is created from the Dwarf web portal, the RPMs must be indexed to create the APT database. This is done by the APT Dwarf module, which uses the genbasedir tool for this purpose. It analyzes RPM packages in a directory tree and builds information files so that that directory tree can be used as a proper APT repository.

The Dwarf database contains information on security (authentication and authorization), repositories types and metadata, mirror repositories, and logging information. Dwarf database contains metadata repository information on buildÕs timestamps, contexts, and descriptions of the repositories, as well as repository types. The rules for creation of mirror repositories are also kept in the Dwarf database. In addition, for security and auditing reasons, the database contains a log of all user-initiated actions. The Dwarf database is realized using the MySQL database technology.

Once the repository is constructed, it is made available by HTTP and FTP servers configured and working on the Dwarf web portal. The DWARF framework provides configurations that should be included in the local HTTP and FTP serversÕ configuration files in order to provide the context of repositories.

%----------------------------------------------------------------------------------------------------%

\section{Monitoring of SEE-GRID infrastructure}
\label{mon}

\begin{figure}[!b]
  \includegraphics[width=0.85\textwidth]{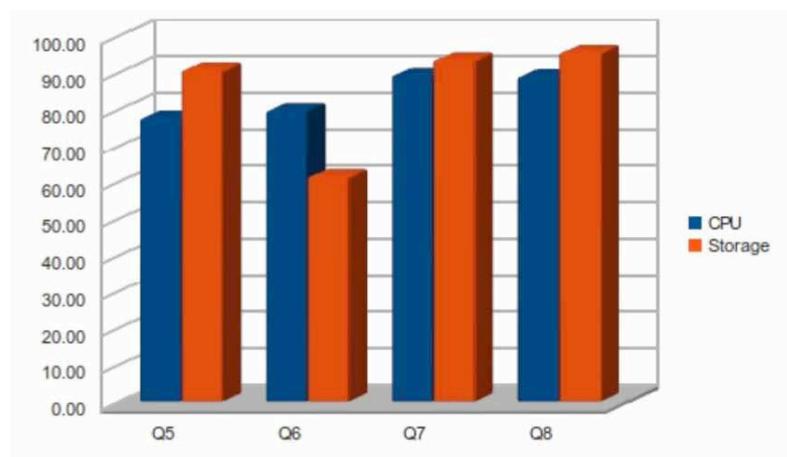}
\caption{Overview of availability of Grid services within the SEE-GRID-SCI infrastructure per quarter in the second (final) year of the project (May 2009 to April 2010).}
\label{avail}
\end{figure}

The monitoring of the heterogeneous and widely geographically dispersed Grid infrastructure is an essential task for achieving the required quality of service to supported user communities. This has been defined through the SEE-GRID Service Level Agreement (SLA), which has served as a prototype for the later adopted EGEE SLA. The monitoring of the performance of sites is not only used for formal assessment of the conformance to SLA, but also for day-to-day Grid operations, since various monitoring tools provide the main channel for identification and diagnostics of operational problems by Grid Operators on Duty and GIMs. The most important such tools are listed in Table~\ref{toolstab}, and we briefly describe them in this section. To illustrate how the conformance of availabilities of Grid services to the adopted SLA was monitored and assessed, Fig.~\ref{avail} gives overview of the availability monitoring results for the second year of the SEE-GRID-SCI project (May 2009 to April 2010). Using the BBmSAM tool (described below), precise measurement of the availability of all services was systematically done, and detailed results were provided at different levels or granularity: per service, per site, per country, and per SEE-GRID infrastructure. For example, the overall availability of resources (weighted by the CPU number of individual clusters) for the last four quarters increased from around 78\% in Q5 (May - July 2009) to around 89\% in Q8 (February - April 2010). Strict enforcement of SLA lead to a steady increase in the availability and reliability of Grid services offered to our target user communities.

\subsection{BBmSAM}
\label{bbmsam}

Availability monitoring of the infrastructure is carried out using the Service Availability Monitoring - SAM \cite{sam} framework developed in EGEE project \cite{egee}, which is further developed and extended by SEE-GRID series of projects and deployed by its SA1 activity. The original SAM system consists of server and client components which communicate using web services. The client initiates periodical tests of the infrastructure and publishes data to the server which stores them in the Oracle database. Main change in the SAM framework in its adaptation for the SEE-GRID community was its porting to MySQL, suitable for the deployment in the region and in line with the SEE-GRID open source philosophy.

BBmSAM \cite{bbmsam,bbmobilesam,standalonesam} Platform is a web application coded in PHP and using the MySQL Database as data storage back-end (although any standard-compliant SQL database server could be used, since it does not rely on any of MySQL-specific features). It has been tested under Apache HTTPD and Microsoft IIS web servers, and should work with any web server supporting PHP (at least through CGI). Main features of BBmSAM system are:
\begin{itemize}
\item{Use of unaltered client and sensor components of EGEE SAM system.}
\item{Synchronization with central HGSM service.}
\item{Use of free and open source technologies.}
\end{itemize}

\begin{figure}[!b]
  \includegraphics[width=0.9\textwidth]{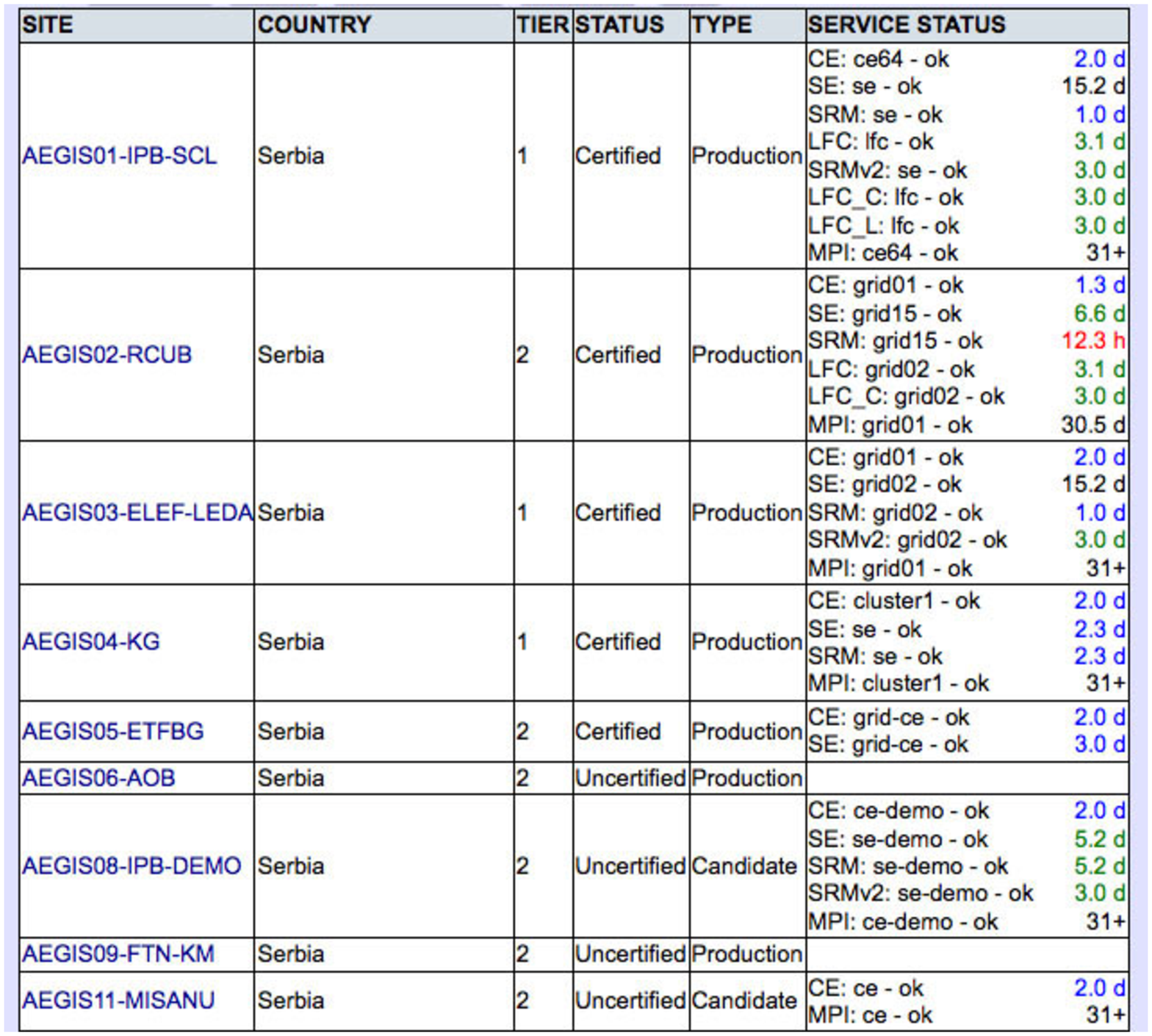}
\caption{Overview of the SEE-GRID BBmSAM web portal.}
\label{bbmsamsnap}
\end{figure}

BBmSAM client and sensors are the same as ones used in the standard EGEE SAM distribution, and they operate in identical way. In designing BBmSAM portal and dependent web services, special care was taken so that the solution would be compatible with EGEE/EGI tools and practices. This was achieved by implementing the same web services in PHP/MySQL implementation as the ones used in the original Java/Oracle-based SAM.

Main part of the BBmSAM web front-end, shown in Fig.~\ref{bbmsamsnap}, is a summary of current results for all tested Grid sites, containing site names, countries and other relevant details for each service.
%--------------------------------------------------%

\subsection{SEE-GRID Accounting Portal}
\label{accounting}

Accounting Portal \cite{accounting} is a web-service based utility to collect and statistically present information on the CPU accounting data for the SEE-GRID computing resources. Its main purpose is to collect and manage accounting data for the sites in SEE-GRID infrastructure. Recently a new publisher was released, capable of collecting and processing data for parallel MPI jobs, which are not properly accounted for when using the standard publisher provided by gLite. The accounting processing structure is based on two services: MPI log parser and accounting publisher. The MPI log parser tool processes PBS Server logs and inserts the data on MPI jobs in the MPI accounting database on the MON node. Afterwards, the accounting publisher aggregates the data from the standard accounting database and MPI database and sends it to the central accounting portal database. The publisher is based on an independent module architecture which allow the two modules (MPI and standard serial) to work independently, so that sites that do not support MPI can use the same publisher.

\begin{figure}[!b]
  \includegraphics[width=0.85\textwidth]{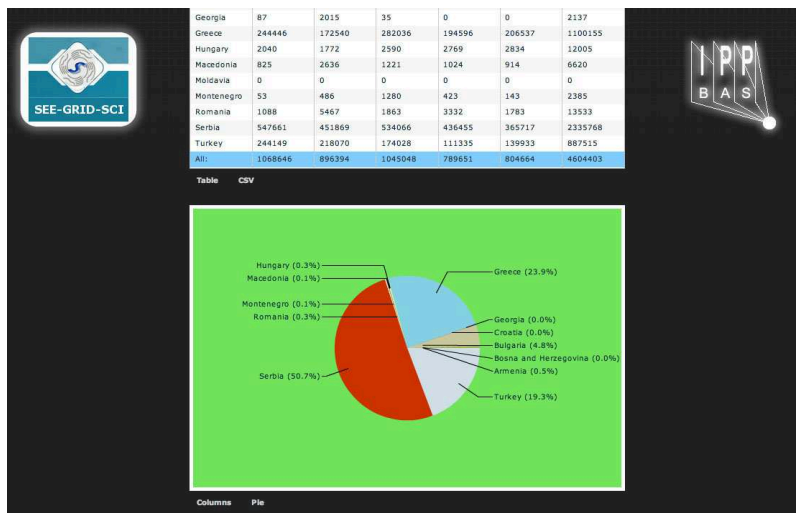}
\caption{Overview of the SEE-GRID accounting portal.}
\label{accountingsnap}
\end{figure}

New web front-end interface of the accounting portal (Fig.~\ref{accountingsnap}) is created to dynamically generates account statistics and charts. It is written in Adobe Flex and Java and implements the MVC design pattern. The View module of the portal is written in Flex, offering an interactive environment with dynamic visualization of the accounting data managed in tables, bar and pie charts. The Interface module is a Java web service which accepts input parameters such as data type, job type, period, rows and columns for the tables.  In addition, it is capable of filtering the data by VO, country and site, offering more flexible data organization. It can also generate SQL queries based on the provided parameters and extract required data from the accounting database. The data are returned in XML form, suitable for import to a variety of other applications. The web portal is hosted on a web server running under Apache Tomcat with installed Apache Axis web-service framework.

%--------------------------------------------------%

\subsection{WatG Browser}
\label{watg}

The What is at the Grid - WatG Browser \cite{watg} is a web-based Grid Information System (GIS) visualization application providing detailed overview of the status and availability of various Grid resources in a given gLite-based e-Infrastructure. It is able to query and present data obtained from Grid information systems at different layers: from local resource information system for a particular Grid service (GRIS), to the Grid site information system (site BDII), and to the top-level information system for the whole Grid infrastructure (top-level BDII). 

\begin{figure}[!b]
  \includegraphics[width=0.95\textwidth]{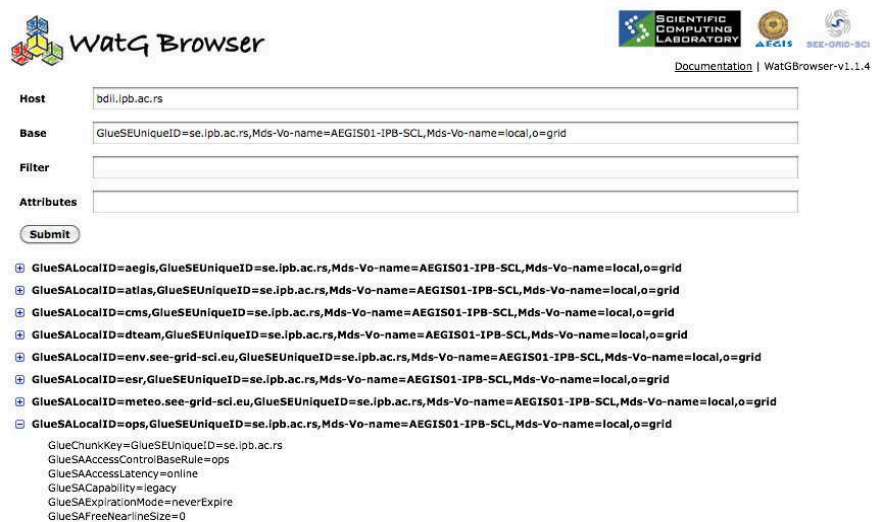}
\caption{Overview of WatG Browser.}
\label{watgsnap}
\end{figure}

The efficient implementation of WatG Browser allows quick and easy navigation through entries and objects of the LDAP tree retrieved by the specified query, even if the size of the output is huge and hierarchically very complex. Highly responsibility is achieved with implementation of partial refreshes and asynchonization of a web page. A partial refresh of WatG application can be observed when an interaction event is triggered, for example click on the plus icon of the LDAP tree. The server processes the information and returns a limited response specific to the data it receives, for example LDAP's subtree that requires given condition. One may notice that WatG server does not send back an entire page, like the conventional "click, wait and refresh" web applications. Instead, WatG client updates the page based on the response. This means that only part of the page is updated. In other words, WatG's initial page (Fig.~\ref{watgsnap}) is treated like a template: WatG server and client exchange the data and the client updates parts of the template based on the data it receives from the server. Another way to think about it is to consider WatG application as driven by events and data, whereas conventional web applications are driven by pages. Asynchronization of the WatG application is reflected in the fact that after sending data to the server, the client can continue processing while the server does its processing in the background. During all this, a user can continue interacting with the client without noticing interruption or a lag in the response. For example, a user can click on any plus or minus icon even during the loading, and in that way a new request will be created and executed afterwards. The client does not have to wait for a response from the server before continuing, as is the case in the traditional, synchronous approach. The WatG Browser is deployed by SCL \cite{scl} and publicly available at the address given in Ref.~\cite{watg}.

%--------------------------------------------------%

\subsection{WMS Monitoring Tool}
\label{wmsmon}

The complex task of computing resources discovery and management on behalf of user applications in the gLite Grid environment is done by the Workload Management System (WMS) service. WMS monitoring tool WMSMon \cite{wmsmon} provides reliable, site-independent, centralized, and uniform monitoring of gLite WMS services.

\begin{figure}[!b]
  \includegraphics[width=0.95\textwidth]{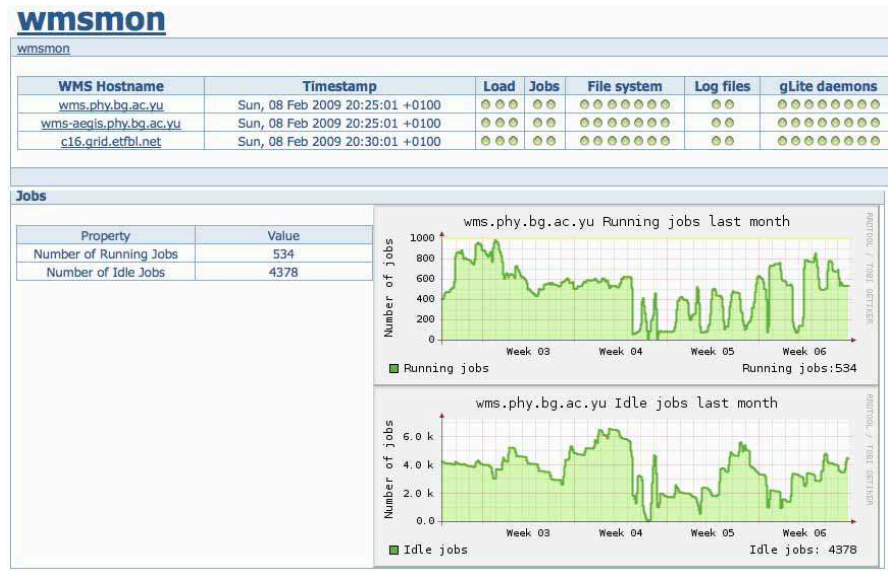}
\caption{Overview of WMSMon Portal.}
\label{wmsmonsnap}
\end{figure}

WMSMon tool, developed and deployed by SCL \cite{scl}, is based on the collector-agent architecture that ensures monitoring of all properties relevant for successful operation of gLite WMS service and triggering of the alarms if certain monitored parameter values exceed predefined limits. In addition, the tool provides links to the appropriate troubleshooting guides when problems are identified.

WMSMon tool consists of two parts of software. The first one, WMSMon Agent, should be installed on all monitored WMS services, and locally aggregates the values of all relevant parameters described in the previous section. The second component of WMSMon software is WMSMon Collector, installed on a specific machine equipped with the web server and gridFTP client, with the aim to collect the data from all WMSMon Agents and to provide web interface to the graphical presentation of the collected data.

WMSMon web portal presents information from diverse WMS sources in a unified way, as can be seen in Fig.~\ref{wmsmonsnap}. The main page provides the aggregated status view of all monitored WMS services from the target Grid infrastructure. This part of the portal presents the data in a simplified way, with the emphasis on WMS services identified not to work properly. The portal also provides links to pages with detailed information and graphs for each monitored WMS service. These pages contain the latest data, as well as historical data presented in the graphical form.

In addition to the main WMSMon instance deployed by SCL \cite{scl}, other instances of WMSMon are installed and used at Grid Operations Centre at CERN \cite{cern} and at NIKHEF \cite{nikhef}.

%----------------------------------------------------------------------------------------------------%

\section{SEE Involvement in High Performance Computing}

The Grid developments in the region, described in this paper, are currently being complemented with supercomputing / High-Performance Computing (HPC) actions. The HP-SEE project \cite{hpsee} (High-Performance Computing Infrastructure for South East EuropeÕs Research Communities) is currently work across several strategic lines of action. First, it is linking the existing HPC facilities in the region into a common infrastructure, and providing operational and management solutions for it. Second, it is striving to open this infrastructure to a wide range of new user communities, including those of non-resourced countries, fostering collaboration and providing advanced capabilities to more researchers, with an emphasis on strategic groups in computational physics, computational chemistry and life sciences. Finally, it acts as a catalyst for establishment of national HPC initiatives, and will act as a SEE bridge for DEISA \cite{deisa}, also presented in this edition, as well as PRACE \cite{prace} infrastructure.

\begin{figure}[!t]
  \includegraphics[width=0.85\textwidth]{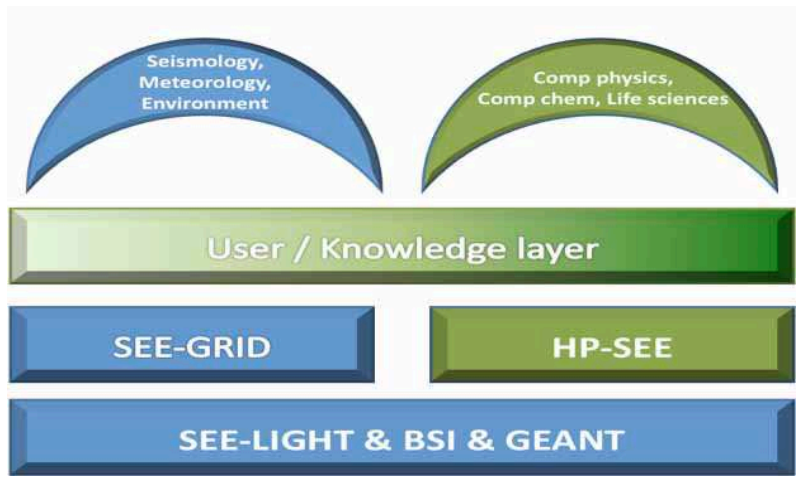}
\caption{SEE eInfrastructure with HPC, and new user communities.}
\label{hpseeoverview}
\end{figure}

Fig.~\ref{hpseeoverview} depicts the multi-dimensional regional eInfrastructure in South-East Europe, where HP-SEE effectively adds the new Research Infrastructure: HPC infrastructure and knowledge / user layer, on top of the existing network plane, and parallel to the existing Grid plane, thus optimising all layers and further enabling a wide range of new cross-border eScience applications to be deployed over the regional eInfrastructure. This approach effectively creates an integrated eInfrastructure for new virtual research communities, and provides a platform for collaboration between ICT engineers and computational scientists dealing with the infrastructure on one hand, and on the other the scientists from diverse scientific communities in the region. 

It should be noted that this vision will provide an integrated infrastructure, where Grid and HPC layers and not mutually exclusive but rather complementary, and tailored for the type of applications supported. Table~\ref{tabhpsee} gives the overview of the current and planned HPC resources that will be available to the HP-SEE Virtual Research Communities within the project.

\begin{table}[!b]
\caption{Current and planned computing power (TFlops) by HP-SEE countries (double precision for CPU and single precision for GPU).}
\label{tabhpsee}
\begin{tabular}{llll}
\hline
Country & 2010 & 2011 & 2012\\
\hline
Greece & 0 & 40 & 80\\
Serbia & 0 &20 &40\\
Bulgaria & 25 & 30+8 GPU & 40+20 GPU\\
Romania & 10 & 20+100 GPU & 30+100 GPU\\
Hungary &1 &30 &60\\
\hline
OVERALL & 36 & 140+108 GPU & 250+120 GPU\\
\hline
\end{tabular}
\end{table}

The available resources will be integrated into a common infrastructure available for the regional Virtual Research Communities. The current and planned HPC infrastructure is heterogeneous, comprising of BlueGene supercomputers, Intel/AMD clusters and enhanced with GPU computing accelerators. Concerning the middleware deployments, we believe the upcoming Unified Middleware Distribution, which will combine Unicore, gLite and ARC will be well suited for the regional HPC infrastructure, taking into account the current situation, where various combinations of these middleware stacks with batch systems and workflow management systems exist. The regional HP-SEE infrastructure will be operated through the operations centre that will be established within the project, which will carry out analysis, requirements capture and evaluation, and deployment of the existing solutions for system management of the regional infrastructure; will identify missing components, and provide optimal solutions. Solutions by system vendors and successful developments from European projects, especially DEISA, and PRACE, will be taken into account. Wherever possible the existing solutions will be adapted and enhanced for deployment in the regional infrastructure. A set of operational tools will be deployed, including user administration, accounting, distributed data management, security, authentication and authorization, monitoring of distributed resources, resource management and allocation, and helpdesk for user support.

The identified target user communities include computational physics, computational chemistry and life sciences. Computational physics is represented by 8 applications from 6 countries and covers the fields of many-body condensed matter physics, including modeling of electron transport, modeling of complex gas dynamics and convection, plasma physics and image processing. Computational chemistry community includes 7 applications from 6 countries, covering the fields of molecular dynamics and simulations, and materials science. Life sciences community has 7 applications from 5 countries, covering the fields of computational biology, computational genomics, computational biophysics and DNA sequencing.

\section{Transition to EGI and Conclusions}
\label{egicon}

Over the period of 6 years and three phases, the SEE-GRID programme aimed at creating independent and sustainable NGIs in each country of the South-Eastern Europe. That has allowed all the countries to participate as full-fledged members of the wider European Grid infrastructure realized through the series of EGEE projects and currently by the European Grid Initiative, EGI \cite{egi}. EGI is established as a coordinating organization for the European Grid Infrastructure, based on the federation of individual NGIs, aiming to support a wide variety of multi-disciplinary user communities. To facilitate the above aim, the SEE-GRID programme has focused both at stimulating the support to policy makers as well as for creating sustainable operational structures in each of the countries in the region.

In particular, on the policy level, the last two years of the SEE-GRID programme have focused on monitoring and improving the status of NGIs in partner countries, and providing support for their evolution and integration into the environment standardized by EGI, aiming to achieve sustainability as active partners in this new pan-European collaboration model. This effort resulted in one of main successes of the project, with all countries of the region currently members or associate members of EGI and participating as partners in the EGI-InSPIRE project.

On the operational level, the focus of SEE-GRID was to create and increase the capacity of Grid resources in the region, create independent and stable operational structures, increase the availability of Grid resources, deploy core services in all countries of the region, as well as to develop geographically distributed network of Grid experts able to provide operational and application level support to end users. At the end of the 6th year of the SEE-GRID programme, all SEE countries are providing such an operational infrastructure for the local and international user communities from the pan-European EGI infrastructure, either as independent NGIs or as a part of the South-Easterrn Europe Regional Operations Centre.

We describe bellow the procedure taken by most of the countries in the region in order to become fully independent operational NGIs from the technical point of view.
The new NGIs use EGIÕs Grid Operations Database, GOCDB \cite{gocdb} to register their NGI management structure, sites and operational personnel.  Most of the SEE NGIs base their operational portal on the central portal that is provided by EGI, performing operations via the NGI view that it offers. In cases like the Greek NGI, a standalone operational portal has been setup. During the course of the SEE-GRID projects the regional Helpdesk was based on OneOrZero as it has also been discussed in Section 4. The SEE-GRID Helpdesk by the end of the SEE-GRID projects was fully integrated with GGUS and therefore it is a candidate system for NGIs, to use as their national Helpdesk solution integrated with the Global Grid User Support, GGUS \cite{ggus}. Further to that Request Tracker, RT \cite{rt}has been integrated with GGUS and can offer the same functionality. Based on the above the NGIs of the regional can select which helpdesk solution to use (either directly GGUS, OneOrZero, or RT). Since infrastructure monitoring in EGI has moved from SAM to Nagios, all new NGIs install and operate their own instance of Nagios that integrates with the rest of EGIs monitoring systems. Finally, SEE NGIs use the Unified Middleware Distribution (UMD) as a central repository for installing basic middleware components while still use the regional repository or even some national repositories, for software packages that are tailored to specific needs of their countries and are not available in UMD.
 
Towards the end of the SEE-GRID-SCI project (May 2010) all NGIs of the project where migrated to EGEE/EGI via the SEE-ROC, utilizing the existing ROC infrastructure and services. Since May 2010 and up to now (January 2011) almost all the NGIÕs have migrated to the EGI operational model. The average time for an NGI to migrate its operational structure from SEE-ROC to EGI is between 1 and 3 months.  

The SEE-GRID programme had pivotal role in bridging the digital divide in the SEE region, in spearheading regional research collaborations, and in creating a strong human network in ICT field paving the way towards full integration of the region into the European Research Area (ERA). This work continues with the HP-SEE project \cite{hpsee}, that aims at bringing together the national HPC infrastructures in the region of South Eastern Europe and the regional Virtual Research Communities of Computational Physics, Computational Chemistry and Life Sciences. Enabling of those user communities to get access to HPC resources for their scientific work is the prime goal of this new project, and demonstrates the success of SEE-GRID series of projects in involving scientists from the region in the development and production use of distributed research infrastructures.

%----------------------------------------------------------------------------------------------------%

\begin{acknowledgements}
This work was supported by the European Commission under FP7 project SEE-GRID-SCI (INFRA-2007-1.2.2. Grant No. 211338), and by the Ministry of Science and Technological Development of the Republic of Serbia (project No. ON141035 and ON171017).
\end{acknowledgements}

\end{document}